\documentclass[a4paper,10pt]{article}
\usepackage{amssymb}
\usepackage{amsmath}
\usepackage{latexsym,amssymb,amsmath}
\usepackage{graphicx}

\DeclareFontFamily{U}{rsf}{} \DeclareFontShape{U}{rsf}{m}{n}{
  <5> <6> rsfs5 <7> <8> <9> rsfs7 <10-> rsfs10}{}
\DeclareMathAlphabet\Scr{U}{rsf}{m}{n} \makeatletter
\@addtoreset{equation}{section} \makeatother

\newcommand{\be}{\begin{equation}}
\newcommand{\ee}{\end{equation}}
\newcommand{\bea}{\begin{eqnarray}}
\newcommand{\eea}{\end{eqnarray}}
\newcommand{\ba}{\begin{array}}
\newcommand{\ea}{\end{array}}
\newcommand{\bit}{\begin{itemize}}
\newcommand{\eit}{\end{itemize}}
\newcommand{\ben}{\begin{enumerate}}
\newcommand{\een}{\end{enumerate}}

\begin{document}


\begin{titlepage}

    \thispagestyle{empty}
    \begin{flushright}
        \hfill{CERN-PH-TH/2010-299}\\
    \end{flushright}

    \vspace{5pt}
    \begin{center}
        { \Huge{\textbf{Charge Orbits and Moduli Spaces\\\vspace{5pt}of Black Hole Attractors}}}\vspace{25pt}
        \vspace{55pt}

        { \Large{\textbf{{Alessio Marrani}}}}\vspace{15pt}

        \large{{\it  Physics Department, Theory Unit, CERN,\\
     CH -1211, Geneva 23, Switzerland\\
     \texttt{Alessio.Marrani@cern.ch}}}

        \vspace{40pt}
        \vspace{20pt} \noindent \large{\textit{Contribution to the Proceedings of
the Workshop\\``Supersymmetry in Mathematics and Physics",\\
February 6--7 2010,\\Department of Mathematics,\\University of
California, Los Angeles, CA, USA}}
\end{center}

\vspace{100pt}

\begin{abstract}
{ \large
 We report on the theory of ``large'' $U$-duality charge
orbits and
related ``moduli spaces'' of extremal black hole attractors in $\mathcal{N}%
=2$, $d=4$ Maxwell-Einstein supergravity theories with symmetric
scalar manifolds, as well as in $\mathcal{N}\geqslant 3$-extended,
$d=4$ supergravities.}
\end{abstract}

\end{titlepage}
\newpage

\section{Introduction\label{Intro}\protect\smallskip}

The \textit{Attractor Mechanism} (AM) \cite{AM-Refs} governs the
dynamics in the scalar manifold of Maxwell-Einstein (super)gravity
theories. It keeps standing as a crucial fascinating key topic
within the international high-energy physics community. Along the
last years, a number of papers have been devoted to the
investigation of attractor configurations of extremal black
$p$-branes in diverse space-time dimensions; for some lists of
Refs., see \textit{e.g.} \cite{Revs}.

The AM is related to dynamical systems with fixed points, describing
the equilibrium
state and the stability features of the system under consideration\footnote{%
We recall that a point $x_{fix}$ where the phase velocity $v\left(
x_{fix}\right) $ vanishes is called a \textit{fixed} point, and it
gives a representation of the considered dynamical system in its
equilibrium state,
\begin{equation*}
v\left( x_{fix}\right) =0.
\end{equation*}
The fixed point is said to be an \textit{attractor} of some motion
$x\left( t\right) $ if
\begin{equation*}
lim_{t\rightarrow \infty }x(t)=x_{fix}.
\end{equation*}
}. When the AM holds, the particular property of the long-range
behavior of the dynamical flows in the considered (dissipative)
system is the
following: in approaching the fixed points, properly named \textit{attractors%
}, the orbits of the dynamical evolution lose all memory of their
initial conditions, but however the overall dynamics remains
completely deterministic.

The first example of AM in supersymmetric systems was discovered in
the theory of static, spherically symmetric, asymptotically flat
extremal dyonic black holes in $\mathcal{N}{=2}$ Maxwell-Einstein
supergravity in $d=4$ and $5$ space-time dimensions (see the first
two Refs. of \cite{AM-Refs}). In the following, we will briefly
present some basic facts about the $d=4$ case.

The multiplet content of a completely general $\mathcal{N}=2$, $d=4$
supergravity theory is the following (see \textit{e.g.}
\cite{N=2-Big}, and Refs. therein):

\begin{enumerate}
\item  the \textit{gravitational} multiplet
\begin{equation}
\left( V_{\mu }^{a},\psi ^{A},\psi _{A},A^{0}\right) ,\label{g-mult}
\end{equation}
described by the \textit{Vielbein} one-form $V^{a}$ ($a=0,1,2,3$)
(together with the spin-connection one-form $\omega ^{ab}$), the
$SU(2)$ doublet of gravitino one-forms $\psi ^{A},\psi _{A}$
($A=1,2$, with the upper and lower
indices respectively denoting right and left chirality, \textit{i.e.} $%
\gamma _{5}\psi _{A}=-\gamma _{5}\psi ^{A}$), and the graviphoton
one-form $A^{0}$;

\item  $n_{V}$ \textit{vector} supermultiplets
\begin{equation}
\left( A^{I},\lambda ^{iA},\overline{\lambda }_{A}^{\overline{i}%
},z^{i}\right) ,\label{v-mult}
\end{equation}
each containing a gauge boson one-form $A^{I}$ ($I=1,...,n_{V}$), a
doublet
of gauginos (zero-form spinors) $\lambda ^{iA},\overline{\lambda }_{A}^{%
\overline{i}}$, and a complex scalar field (zero-form) $z^{i}$ ($%
i=1,...,n_{V}$). The scalar fields $z^{i}$ can be regarded as
coordinates on
a complex manifold $\mathcal{M}_{n_{V}}$ ($dim_{\mathbb{C}}\mathcal{M}%
_{n_{V}}=n_{V}$), which is actually a \textit{special K\"{a}hler}
manifold;

\item  $n_{H}$ \textit{hypermultiplets}
\begin{equation}
\left( \zeta _{\alpha },\zeta ^{\alpha },q^{u}\right)
,\label{h-mult}
\end{equation}
each formed by a doublet of zero-form spinors, that is the hyperinos
$\zeta _{\alpha },\zeta ^{\alpha }$ ($\alpha =1,...,2n_{H}$), and
four real scalar fields $q^{u}$ ($u=1,...,4n_{H}$), which can be
considered as coordinates of
a quaternionic manifold $\mathcal{Q}_{n_{H}}$ ($dim_{\mathbb{H}}\mathcal{Q}%
_{n_{H}}=n_{H}$).
\end{enumerate}

\textit{At least} in absence of gauging, the $n_{H}$ hypermultiplets
are spectators in the AM. This can be understood by looking at the
transformation properties of the Fermi fields: the hyperinos $\zeta
_{\alpha },\zeta ^{\alpha }$'s transform independently on the vector
fields, whereas the gauginos' supersymmetry transformations depend
on the Maxwell vector fields. Consequently, the contribution of the
hypermultiplets can be dynamically decoupled from the rest of the
physical system; in particular, it is also completely independent
from the evolution dynamics of the complex scalars $z^{i}$'s coming
from the vector multiplets (\textit{i.e.} from the evolution flow in
$\mathcal{M}_{n_{V}}$). By disregarding for simplicity's sake the
fermionic and gauging terms, the supersymmetry transformations of
hyperinos read (see \textit{e.g.} \cite{N=2-Big}, and Refs. therein)
\begin{equation}
\delta \zeta _{\alpha }=i\mathcal{U}_{u}^{B\beta }\partial _{\mu
}q^{u}\gamma ^{\mu }\varepsilon ^{A}\epsilon _{AB}\mathbb{C}_{\alpha
\beta }, \label{hyperinos}
\end{equation}
implying the asymptotical configurations of the quaternionic scalars
of the hypermultiplets to be unconstrained, and therefore to vary
continuously in the manifold $\mathcal{Q}_{n_{H}}$ of the related
quaternionic non-linear sigma model.

Thus, as far as ungauged theories are concerned, for the treatment of AM one can restrict to consider $%
\mathcal{N}=2$, $d=4$ Maxwell-Einstein supergravity, in which
$n_{V}$ vector
multiplets (\ref{v-mult}) are coupled to the gravity multiplet (\ref{g-mult}%
). The relevant dynamical system to be considered is the one related
to the radial evolution of the configurations of complex scalar
fields of such $n_{V}$ vector multiplets. When approaching the event
horizon of the black hole, the scalars dynamically run into fixed
points, taking values which are only function (of the ratios) of the
electric and magnetic charges associated to Abelian Maxwell vector
potentials under consideration.

The inverse distance to the event horizon is the fundamental
evolution parameter
in the dynamics towards the fixed points represented by the \textit{%
attractor }configurations of the scalar fields. Such near-horizon
configurations, which ``attracts'' the dynamical evolutive flows in
$\mathcal{M}_{n_{V}}$, are completely independent on the initial
data of such an evolution, \textit{i.e.} on the spatial asymptotical
configurations of the scalars. Consequently, for what concerns the
scalar dynamics, the system completely loses memory of its initial
data, because the dynamical evolution is ``attracted'' by some fixed
configuration points, purely depending on the electric and magnetic
charges.

Recently, intriguing connections with the (quantum) theory of
information arose out \cite{QIT-Refs}.

In the framework of supergravity theories, extremal black holes can
be interpreted as BPS (Bogomol'ny-Prasad-Sommerfeld)-saturated
\cite{BPS} interpolating metric singularities in the low-energy
effective limit of higher-dimensional superstrings or $M$-theory
\cite{GT-p-branes}. Their asymptotically relevant parameters include
the ADM mass \cite{ADM}, the electrical and magnetic charges
(defined by integrating the fluxes of related field strengths over
the $2$-sphere at infinity), and the asymptotical values of the
(dynamically relevant set of) scalar fields. The AM implies that the
class of black holes under consideration loses all its ``scalar
hair'' within the near-horizon geometry. This means that the
extremal black hole solutions, in the near-horizon limit in which
they approach the Bertotti-Robinson $AdS_{2}\times S^{2}$
conformally flat metric \cite{BR}, are characterized only by
electric and magnetic charges, but not by the continuously-varying
asymptotical values of the scalar fields.

An important progress in the geometric interpretation of the AM was
achieved in the last Ref. of \cite{AM-Refs}, in which the attractor
near-horizon scalar configurations were related to the critical
points of a suitably defined black hole effective potential function
$V_{BH}$, whose explicit form in maximal supergravity is
\textit{e.g.} given by Eq. (\ref{V_BH-N=8}) below. In general,
$V_{BH}$ is a positive definite function of scalar fields and
electric and magnetic charges, and its non-degenerate critical
points in $\mathcal{M}_{n_{V}}$
\begin{equation}
\forall i=1,...,n_{V},\frac{\partial V_{BH}}{\partial
z^{i}}=0:~\left. V_{BH}\right| _{\frac{\partial V_{BH}}{\partial
z}=0}>0,\label{crit-points}
\end{equation}
fix the scalar fields to depend only on electric and magnetic fluxes
(charges). In the Einstein two-derivative approximation, the
(semi)classical Bekenstein-Hawking entropy ($S_{BH}$) - area
($A_{H}$) formula \cite{BH} yields the (purely charge-dependent)
black hole entropy $S_{BH}$ to be
\begin{equation}
S_{BH}=\pi \frac{A_{H}}{4}=\pi \left. V_{BH}\right| _{\frac{\partial V_{BH}}{%
\partial z}=0}=\pi \sqrt{\left| \mathcal{I}_{4}\right| },\label%
{BH-entropy-area-formula}
\end{equation}
where $\mathcal{I}_{4}$ is the unique independent invariant
homogeneous
polynomial (quartic in charges) in the relevant representation $\mathbf{R}%
_{V}$ of $G$ in which the charges sit (see Eq. (\ref{emb}) and
discussion
below). The last step of (\ref{BH-entropy-area-formula}) does not apply to $%
d=4$ supergravity theories with quadratic charge polynomial
invariant, namely to the $\mathcal{N}=2$ \textit{minimally coupled}
sequence \cite
{Luciani} and to the $\mathcal{N}=3$ \cite{N=3} theory; in these cases, in (%
\ref{BH-entropy-area-formula}) $\sqrt{\left| \mathcal{I}_{4}\right|
}$ gets replaced by $\left| \mathcal{I}_{2}\right| $.

In presence of $n=n_{V}+1$ Abelian vector fields, the fluxes sit in
a $2n$-dimensional representation $\mathbf{R}_{V}$ of the $U$-duality group $%
G$, defining the embedding of $G$ itself into $Sp\left(
2n,\mathbb{R}\right) $, which is the largest group acting linearly
on the fluxes themselves:
\begin{equation}
G\overset{\mathbf{R}_{V}}{\subsetneq }Sp\left( 2n,\mathbb{R}\right)
.\label{emb}
\end{equation}
It should be pointed out that we here refer to $U$-duality as the
continuous version of the $U$-duality groups introduced in
\cite{HT}. This is consistent with the assumed (semi-)classical
limit of large charges,
also indicated by the fact that we consider $Sp\left( 2n,\mathbb{R}\right) $%
, and not $Sp\left( 2n,\mathbb{Z}\right) $ (no
Dirac-Schwinger-Zwanziger quantization condition is implemented on
the fluxes themselves).

After \cite{FG1,DFL,FM}, the the $\mathbf{R}_{V}$-representation
space of the $U$-duality group is known to exhibit a stratification
into disjoint classes of orbits, which can be defined through
invariant sets of constraints on the (lowest order, actually unique)
$G$-invariant $\mathcal{I} $ built out of the symplectic
representation $\mathbf{R}_{V}$. It is here worth remarking the
crucial distinction between the ``large'' orbits and ``small''
orbits. While the former have $\mathcal{I}\neq 0$ and support an
attractor behavior of the scalar flow in the near-horizon geometry
of the extremal black hole background\textit{\ }\cite{AM-Refs}, for
the latter the Attractor Mechanism does not hold, they have
$\mathcal{I}=0$ and thus they
correspond to solutions with vanishing Bekenstein-Hawking \cite{BH} entropy (%
\textit{at least} at the Einsteinian two-derivative
level).\smallskip

This short report, contributing to the Proceedings of the Workshop \textit{%
``Supersymmetry in Mathematics and Physics'' }(organized by Prof. R.
Fioresi and Prof. V. S. Varadarajan), held on February 2010 at the
Department of Mathematics of the University of California at Los
Angeles, presents the main \ results of the theory of $U$-duality
charge orbits and ``moduli spaces'' of extremal black hole attractor
solutions in supergravity theories with $\mathcal{N}\geqslant 2$
supercharges in $d=4$ space-time dimensions. In particular,
$\mathcal{N}=2$ Maxwell-Einstein theories with symmetric scalar
manifolds will be considered.\medskip

The plan of this short review is as follows.

Sec. \ref{Sect2} introduces the ``large'' (\textit{i.e.}
attractor-supporting) charge orbits of the $\mathcal{N}=2$, $d=4$ \textit{%
symmetric} Maxwell-Einstein supergravities, namely of those $\mathcal{N}=2$
supergravity theories in which a certain number of Abelian vector multiplets
is coupled to the gravity multiplet, and the corresponding complex scalars
span a special K\"{a}hler manifold which is also a symmetric coset $\frac{G}{%
H_{0}\times U\left( 1\right) }$, where $G$ is the $U$-duality group and $%
H_{0}\times U\left( 1\right) $ is its maximal compact subgroup.

Then, Sec. \ref{Sect3} is devoted to the analysis of the ``large''
charge orbits of the maximal $\mathcal{N}=8$ supergravity theory.
The non-compactness of the stabilizer groups of such (generally
non-symmetric) coset orbits gives rise to the so-called ``moduli
spaces'' of attractor solutions, namely proper subspaces of the
scalar manifold of the theory in which the Attractor Mechanism is
not active.

The ``moduli spaces'' of the various classes of non-supersymmetric
attractors in $\mathcal{N}=2$, $d=4$ \textit{symmetric}
Maxwell-Einstein supergravities are then reported and discussed in
Sec. \ref{Sect4}.

The short Sec. \ref{N>2,d=4} concludes the paper, analyzing the
attractor-supporting orbits of $\mathcal{N}\geqslant 3$-extended
``pure'' and matter-coupled theories, whose scalar manifolds are all
symmetric.

\section{Charge Orbits of $\mathcal{N}=2$, $d=4$\\Symmetric Maxwell-Einstein
Supergravities\label{Sect2}}

$\mathcal{N}=2$, $d=4$ Maxwell-Einstein supergravity theories \cite{GST}
with homogeneous symmetric special K\"{a}hler vector multiplets' scalar
manifolds $\frac{G}{H_{0}\times U(1)}$ will be shortly referred to as
\textit{symmetric} Maxwell-Einstein supergravities. The various symmetric
non-compact special K\"{a}hler spaces $\frac{G}{H_{0}\times U(1)}$ (with $%
H_{0}\times U\left( 1\right) $ being the maximal compact subgroup with
symmetric embedding (\textit{mcs}) of $G$, the $d=4$ $U$-duality group) have
been classified in \cite{CVP,dWVVP} (see \textit{e.g.} \cite{LA08} for a
recent account), and they are reported in Table 1.

\begin{table}[h!]
\par
\begin{center}
\begin{tabular}{|c||c|c|c|}
\hline & $
\begin{array}{c}
\\
\frac{G}{H_{0}\times U(1)} \\
~
\end{array}
$ & $
\begin{array}{c}
\\
r \\
~
\end{array}
$ & $
\begin{array}{c}
\\
$dim$_{\mathbb{C}}\equiv n_{V} \\
~
\end{array}
$ \\ \hline\hline $
\begin{array}{c}
\textit{minimal~coupling} \\
n\in \mathbb{N}
\end{array}
$ & $\mathbb{CP}^{n}\equiv \frac{SU(1,n)}{U(1)\times SU(n)}~$ & $1$
& $n~$ \\ \hline $
\begin{array}{c}
\\
\mathbb{R}\oplus \mathbf{\Gamma }_{1,n-1},~n\in \mathbb{N} \\
~
\end{array}
$ & $\frac{SL(2,\mathbb{R})}{SO\left( 2\right) }\times
\frac{SO(2,n)}{SO(2)\times SO(n)}~$ & $
\begin{array}{c}
\\
2~(n=1) \\
3~(n\geqslant 2)
\end{array}
~$ & $n+1$ \\ \hline $
\begin{array}{c}
\\
J_{3}^{\mathbb{O}} \\
~
\end{array}
$ & $\frac{E_{7(-25)}}{E_{6(-78)}\times U(1)}$ & $3$ & $27$ \\
\hline $
\begin{array}{c}
\\
J_{3}^{\mathbb{H}} \\
~
\end{array}
$ & $\frac{SO^{\ast }(12)}{U(6)}~$ & $3$ & $15$ \\ \hline $
\begin{array}{c}
\\
J_{3}^{\mathbb{C}} \\
~
\end{array}
$ & $\frac{SU(3,3)}{S\left( U(3)\times U(3)\right) }$ & $3$ & $9~$
\\ \hline $
\begin{array}{c}
\\
J_{3}^{\mathbb{R}} \\
~
\end{array}
$ & $\frac{Sp(6,\mathbb{R})}{U(3)}$ & $3$ & $6$ \\ \hline
\end{tabular}
\end{center}
\caption{\textbf{Riemannian globally symmetric} \textbf{non-compact
special K\"{a}hler spaces (\textit{alias} vector multiplets' scalar
manifolds of the \textit{symmetric} }$\mathcal{N}=2$\textbf{,}
$d=4$\textbf{\ Maxwell Einstein supergravity theories).
}$r$\textbf{\ denotes the rank of the manifold, whereas
}$n_{V}$\textbf{\ stands for the number of vector multiplets}}
\end{table}

\begin{table}[h!]
\begin{center}
\begin{tabular}{|c||c|c|c|}
\hline & $
\begin{array}{c}
\\
\frac{1}{2}\text{-BPS orbit } \\
~~\mathcal{O}_{\frac{1}{2}-BPS}=\frac{G}{H_{0}} \\
~
\end{array}
$ & $
\begin{array}{c}
\\
\text{nBPS }Z_{H}\neq 0\text{ orbit} \\
\mathcal{O}_{nBPS,Z_{H}\neq 0}=\frac{G}{\widehat{H}}~ \\
~
\end{array}
$ & $
\begin{array}{c}
\\
\text{nBPS }Z_{H}=0\text{ orbit} \\
\mathcal{O}_{nBPS,Z_{H}=0}=\frac{G}{\widetilde{H}}~ \\
~
\end{array}
$ \\ \hline\hline $
\begin{array}{c}
\\
\mathit{minimal~coupling} \\
~n\in \mathbb{N}
\end{array}
$ & $\frac{SU(1,n)}{SU(n)}~$ & $-$ & $\frac{SU(1,n)}{SU(1,n-1)}~$ \\
\hline $
\begin{array}{c}
\\
\mathbb{R}\oplus \mathbf{\Gamma }_{1,n-1} \\
~n\in \mathbb{N}
\end{array}
$ & $\frac{SL(2,\mathbb{R})}{SO\left( 2\right) }\times \frac{SO(2,n)}{SO(n)}~$ & $%
\frac{SL(2,\mathbb{R})}{SO\left( 1,1\right) }\times \frac{SO(2,n)}{SO(1,n-1)}~$ & $%
\frac{SL(2,\mathbb{R})}{SO\left( 2\right) }\times
\frac{SO(2,n)}{SO(2,n-2)}$
\\ \hline $
\begin{array}{c}
\\
J_{3}^{\mathbb{O}} \\
~
\end{array}
$ & $\frac{E_{7(-25)}}{E_{6}}$ & $\frac{E_{7(-25)}}{E_{6(-26)}}$ & $\frac{%
E_{7(-25)}}{E_{6(-14)}}~$ \\ \hline $
\begin{array}{c}
\\
J_{3}^{\mathbb{H}} \\
~
\end{array}
$ & $\frac{SO^{\ast }(12)}{SU(6)}~$ & $\frac{SO^{\ast
}(12)}{SU^{\ast }(6)}~$ & $\frac{SO^{\ast }(12)}{SU(4,2)}~$ \\
\hline $
\begin{array}{c}
\\
J_{3}^{\mathbb{C}} \\
~
\end{array}
$ & $\frac{SU(3,3)}{SU(3)\times SU(3)}$ &
$\frac{SU(3,3)}{SL(3,\mathbb{C})}$ & $\frac{SU(3,3)}{SU(2,1)\times
SU(1,2)}~$ \\ \hline $
\begin{array}{c}
\\
J_{3}^{\mathbb{R}} \\
~
\end{array}
$ & $\frac{Sp(6,\mathbb{R})}{SU(3)}$ & $\frac{Sp(6,\mathbb{R})}{SL(3,\mathbb{%
R})}$ & $\frac{Sp(6,\mathbb{R})}{SU(2,1)}$ \\ \hline
\end{tabular}
\end{center}
\caption{\textbf{Charge orbits of attractors in \textit{symmetric }}$%
\mathcal{N}$\textbf{$=2$, $d=4$ Maxwell-Einstein supergravities}}
\end{table}

All these theories can be obtained by dimensional reduction of the minimal $%
\mathcal{N}=2$, $d=5$ supergravities \cite{GST}, and they all have cubic
prepotential holomorphic functions. The unique exception is provided by the
theories with $\mathbb{CP}^{n}$ scalar manifolds, describing the \textit{%
minimal coupling} of $n$ Abelian vector multiplets to the gravity multiplet
itself \cite{Luciani} (see also \cite{BFGM1,Gnecchi-1}); in this case, the
prepotential is quadratic in the scalar fields,and thus $C_{ijk}=0$.

By disregarding the $\mathbb{CP}^{n}$ sequence, the cubic prepotential of
all these theories is related to the norm form of the Euclidean degree-$3$
Jordan algebra that defines them \cite{GST}. The reducible sequence in the
third row of Table 1, usually referred to as the \textit{generic Jordan
family}, is based on the sequence of \textit{reducible} Euclidean Jordan
algebras $\mathbb{R}\oplus \mathbf{\Gamma }_{1,n-1}$, where $\mathbb{R}$
denotes the $1$-dimensional Jordan algebra and $\mathbf{\Gamma }_{1,n-1}$
stands for the degree-$2$ Jordan algebra with a quadratic form of Lorentzian
signature $\left( 1,n-1\right) $, which is nothing but the Clifford algebra
of $O\left( 1,n-1\right) $ \cite{Jordan}.

\textbf{\ }Then, four other theories exist, defined by the irreducible
degree-$3$ Jordan algebras $J_{3}^{\mathbb{O}}$, $J_{3}^{\mathbb{H}}$, $%
J_{3}^{\mathbb{C}}$ and $J_{3}^{\mathbb{R}}$, namely the Jordan algebras of
Hermitian $3\times 3$ matrices over the four division algebras $\mathbb{O}$
(octonions), $\mathbb{H}$ (quaternions), $\mathbb{C}$ (complex numbers) and $%
\mathbb{R}$ (real numbers) \cite{GST,Jordan,Jacobson,Guna1,GPR}. Because of
their symmetry groups fit in the celebrated \textit{Magic Square} of
Freudenthal, Rozenfeld and Tits \cite{Freudenthal2,magic}, these theories
have been named \textit{``magic''}. By defining $A\equiv $dim$_{\mathbb{R}}%
\mathbb{A}$ ($=8,4,2,1$ for $\mathbb{A}=\mathbb{O},\mathbb{H},\mathbb{C},%
\mathbb{R}$, respectively), the complex dimension of the scalar manifolds of
the ``magic'' Maxwell-Einstein theories is $3\left( A+1\right) $. It should
also be recalled that the $\mathcal{N}=2$ ``magic'' theory based on $J_{3}^{%
\mathbb{H}}$ shares the same bosonic sector with the $\mathcal{N}=6$
``pure'' supergravity (see \textit{e.g.} \cite
{ADF-U-duality-revisited,Ferrara-Gimon,Samtleben-twin}), and accordingly in
this case the attractors enjoy a ``dual'' interpretation \cite{BFGM1}.
Furthermore, it should also be remarked that $J_{2}^{\mathbb{A}}\sim \mathbf{%
\Gamma }_{1,A+1}$ (see \textit{e.g.} the eighth Ref. of \cite{Revs}).

Within these theories, the ``large''charge orbits, \textit{i.e.} the ones
supporting extremal black hole attractors have a non-maximal (nor generally
symmetric) coset structure. The results \cite{BFGM1} are reported in Table
2. After \cite{FG1}, the charge orbit supporting ($\frac{1}{2}$-)BPS
attractors has coset structure
\begin{equation}
\mathcal{O}_{BPS}=\frac{G}{H_{0}}\text{,~with~}H_{0}\times U(1)\overset{mcs}{%
\subsetneq }G.  \label{O_BPS-N=2}
\end{equation}
As shown in \cite{BFGM1}, there are other two charge orbits supporting
extremal black hole attractors, and they are both non-supersymmetric (not
saturating the BPS bound \cite{BPS}). One has non-vanishing $\mathcal{N}=2$
central charge at the horizon ($Z_{H}\neq 0$), with coset structure
\begin{equation}
\mathcal{O}_{nBPS,Z_{H}\neq 0}=\frac{G}{\widehat{H}}\text{,~with~}\widehat{H}%
\times SO\left( 1,1\right) \subsetneq G,  \label{O_nBPS-Z<>0}
\end{equation}
where $\widehat{H}$ denotes the $d=5$ $U$-duality group, and thus $SO\left(
1,1\right) $ corresponds to the $S^{1}$-radius in the Kaluza-Klein reduction
$d=5\rightarrow 4$. Also the remaining attractor-supporting charge orbit is
non-supersymmetric, but it corresponds to $Z_{H}=0$; its coset structure
reads
\begin{equation}
\mathcal{O}_{nBPS,Z_{H}=0}=\frac{G}{\widetilde{H}}\text{,~with~}\widetilde{H}%
\times U\left( 1\right) \subsetneq G.  \label{O_nBPS-Z=0}
\end{equation}
It is worth remarking that $\widehat{H}$ and $\widetilde{H}$ are the only
two non-compact forms of $H_{0}$ such that the group embedding in the
right-hand side of (\ref{O_nBPS-Z=0}) and (\ref{O_nBPS-Z<>0}) are both
maximal and symmetric (see \textit{e.g.} \cite{Gilmore,Helgason,Slansky}).

Due to (\ref{O_BPS-N=2}), $H_{0}$ is the maximal compact symmetry group of
the particular class of non-degenerate critical points of the effective
black hole potential $V_{BH}$ corresponding to BPS attractors. On the other
hand, the maximal compact symmetry group of the non-BPS $Z_{H}\neq 0$ and
non-BPS $Z_{H}=0$ critical points of $V_{BH}$ respectively is
\begin{equation}
\widehat{h}=\text{\textit{mcs}}\left( \widehat{H}\right) ;~~\widetilde{h}=%
\text{\textit{mcs}}\left( \widetilde{H}\right) .  \label{N=2-mcs's}
\end{equation}
Actually, in the non-BPS $Z_{H}=0$ case, the maximal compact symmetry is $%
\widetilde{h}^{\prime }\equiv \frac{\widetilde{h}}{U(1)}$; see e.g. \cite
{BFGM1} for further details.

General results on the rank $\frak{r}$ of the $2n_{V}\times 2n_{V}$ Hessian
matrix $\mathbf{H}$ of $V_{BH}$ are known. Firstly, the BPS (non-degenerate)
critical points of $V_{BH,\mathcal{N}=2}$ are stable, and thus $\mathbf{H}%
_{BPS}$ has no massless modes (see the fifth Ref. of \cite{AM-Refs}), and
its rank is maximal: $\frak{r}_{BPS}=2n_{V}$. Furthermore, the analysis of
\cite{BFGM1} showed that for the other two classes of (non-degenerate)
non-supersymmetric critical points of $V_{BH,\mathcal{N}=2}$, the rank of $%
\mathbf{H}$ is model-dependent:
\begin{eqnarray}
\mathbb{CP}^{n} &:&\frak{r}_{nBPS,Z_{H}=0}=2; \\
\mathbb{R}\oplus \mathbf{\Gamma }_{1,n-1} &:&\left\{
\begin{array}{l}
\frak{r}_{nBPS,Z_{H}\neq 0}=n+2; \\
\\
\frak{r}_{nBPS,Z_{H}=0}=6;
\end{array}
\right.  \\
J_{3}^{\mathbb{A}} &:&\left\{
\begin{array}{l}
\frak{r}_{nBPS,Z_{H}\neq 0}=3A+4; \\
\\
\frak{r}_{nBPS,Z_{H}=0}=2A+6.
\end{array}
\right.
\end{eqnarray}

\section{$\mathcal{N}=8$, $d=4$ Supergravity \label{Sect3}}

The analysis of extremal black hole attractors in the theory with the
maximal number of supercharges, namely in $\mathcal{N}=8$, $d=4$
supergravity, provides a simpler, warm-up framework for the analysis and
classification of the ``moduli spaces'' of the two classes ($%
Z_{H}\neq 0$ and $Z_{H}=0$) of non-BPS attractors of quarter-minimal
Maxwell-Einstein supergravities with symmetric scalar manifolds, which have
been introduced in Sec. \ref{Sect2}.

Maximal supergravity in four dimensions is based on the real, rank-$7$, $70$%
-dimensional homogeneous symmetric manifold
\begin{equation}
\frac{G_{\mathcal{N}=8}}{H_{\mathcal{N}=8}}=\frac{E_{7(7)}}{SU(8)},
\end{equation}
where $SU\left( 8\right) =mcs\left( E_{7\left( 7\right) }\right) $. After
\cite{FG1,DFL,FM,FK-N=8,Ferrara-Marrani-1}, two classes of (non-degenerate)
critical points of $V_{BH,\mathcal{N}=8}$ are known to exist:

\begin{itemize}
\item  the $\frac{1}{8}$-BPS class, supported by the orbit
\begin{equation}
\mathcal{O}_{\frac{1}{8}-BPS,\mathcal{N}=8}\equiv \frac{G_{\mathcal{N}=8}}{%
\mathcal{H}_{\mathcal{N}=8}}=\frac{E_{7(7)}}{E_{6(2)}}\text{,~}%
E_{6(2)}\times U\left( 1\right) \subsetneq E_{7\left( 7\right) };
\label{N=8-BPS-large}
\end{equation}

\item  the non-BPS class, supported by the orbit
\begin{equation}
\mathcal{O}_{nBPS,\mathcal{N}=8}\equiv \frac{G_{\mathcal{N}=8}}{\widehat{%
\mathcal{H}}_{\mathcal{N}=8}}=\frac{E_{7(7)}}{E_{6(6)}}\text{,~}%
E_{6(6)}\times SO\left( 1,1\right) \subsetneq E_{7\left( 7\right) }.
\label{N=8-nBPS}
\end{equation}
\end{itemize}

Both charge orbits $\mathcal{O}_{\frac{1}{8}-BPS,\mathcal{N}=8}$ and $%
\mathcal{O}_{nBPS,\mathcal{N}=8}$ belong to the fundamental representation
space $\mathbf{56}$ of the maximally non-compact (split) form $E_{7\left(
7\right) }$ of the exceptional group $E_{7}$. The embeddings in the
right-hand side of Eqs. (\ref{N=8-BPS-large}) and (\ref{N=8-nBPS}) are both
maximal and symmetric (see \textit{e.g.} \cite{Gilmore,Slansky}). Among all
non-compact forms of the exceptional Lie group $E_{6}$ (\textit{i.e. }$%
E_{6\left( -26\right) }$, $E_{6\left( -14\right) }$, $E_{6\left( 2\right) }$
and $E_{6(6)}$), $E_{6(2)}$ and $E_{6(6)}$ are the only two which are
maximally and symmetrically embedded (through an extra group factor $U\left(
1\right) $ or $SO\left( 1,1\right) $) into $E_{7\left( 7\right) }$.

In the maximal theory, the Hessian matrix $\mathbf{H}_{\mathcal{N}=8}$ of
the effective potential $V_{BH,\mathcal{N}=8}$ is a square $70\times 70$
symmetric matrix. At $\frac{1}{8}$-BPS attractor points, $\mathbf{H}_{%
\mathcal{N}=8}$ has rank $30$, with $40$ massless modes \cite
{ADF-U-duality-d=4} sitting in the representation $\left( \mathbf{20},%
\mathbf{2}\right) $ of the enhanced $\frac{1}{8}$-BPS symmetry group $%
SU(6)\times SU(2)=mcs\left( \mathcal{H}_{\mathcal{N}=8}\right) $ \cite
{Ferrara-Marrani-1}. Moreover, at non-BPS attractor points, $\mathbf{H}_{%
\mathcal{N}=8}$ has rank $28$, with $42$ massless modes sitting in the
representation $\mathbf{42}$ of the enhanced non-BPS symmetry group $%
USp(8)=mcs\left( \widehat{\mathcal{H}}_{\mathcal{N}=8}\right) $ \cite
{Ferrara-Marrani-1}. Actually, the massless modes of $\mathbf{H}_{\mathcal{N}%
=8}$ are ``flat'' directions of $V_{BH,\mathcal{N}=8}$ at the
corresponding classes of its critical points. Thus, such ``flat''
directions of the critical $V_{BH,\mathcal{N}=8}$ span some ``moduli
spaces'' of the attractor solutions \cite{Ferrara-Marrani-2},
corresponding to the scalar degrees of freedom which are not
stabilized by the \textit{Attractor Mechanism} \cite{AM-Refs} at the
black hole event horizon. In the $\mathcal{N}=8$ case, such ``moduli
spaces'' are the following real symmetric sub-manifolds of
$\frac{E_{7(7)}}{SU(8)}$ itself \cite {Ferrara-Marrani-2}:
\begin{eqnarray}
\frac{1}{8}\text{-BPS} &:&\mathcal{M}_{\frac{1}{8}-BPS}=\frac{\mathcal{H}_{%
\mathcal{N}=8}}{mcs\left( \mathcal{H}_{\mathcal{N}=8}\right) }=\frac{E_{6(2)}%
}{SU(6)\times SU(2)},~\text{dim}_{\mathbb{R}}=40\text{,~rank}=4;
\label{N=8-BPS-large-mod} \\
&&  \notag \\
\text{non-BPS} &:&\mathcal{M}_{nBPS}=\frac{\widehat{\mathcal{H}}_{\mathcal{N}%
=8}}{mcs\left( \widehat{\mathcal{H}}_{\mathcal{N}=8}\right) }=\frac{E_{6(6)}%
}{USp(8)},~\text{dim}_{\mathbb{R}}=42\text{,~rank}=6.  \label{N=8-nBPS-mod}
\end{eqnarray}
It is easy to realize that $\mathcal{M}_{\frac{1}{8}-BPS}$ and $\mathcal{M}%
_{nBPS}$ are nothing but the cosets of the non-compact stabilizer of
the corresponding supporting charge orbit ($E_{6(2)}$ and
$E_{6(6)}$, respectively) and of its \textit{mcs}. Actually, this is
the very structure of all ``moduli spaces'' of attractors (see
Sects. \ref{Sect4} and \ref {N>2,d=4}). Moreover,
$\mathcal{M}_{nBPS}$ is nothing but the scalar manifold of
$\mathcal{N}=8$, $d=5$ supergravity. This holds more in general,
and, as given by the treatment of Sec. \ref{Sect4} (see also Table
3), the ``moduli space'' of $\mathcal{N}=2$, $d=4$ non-BPS
$Z_{H}\neq 0$ attractors is nothing but the scalar manifold of the
$d=5$ uplift of the corresponding theory \cite{Ferrara-Marrani-2}
(see also \cite{CFM1}).

Following \cite{Ferrara-Marrani-2} and considering the maximal supergravity
theory, we now explain the reason why the ``flat'' directions of the Hessian
matrix of the effective potential at its critical points actually span a
``moduli space'' (for a recent discussion, see also \cite{ADFT-FO-1}).

Let us start by recalling that $V_{BH,\mathcal{N}=8}$ is defined as
\begin{equation}
V_{BH,\mathcal{N}=8}\equiv \frac{1}{2}Z_{AB}\left( \phi ,Q\right) \overline{Z%
}^{AB}\left( \phi ,Q\right) ,  \label{V_BH-N=8}
\end{equation}
where $Z_{AB}$ is the antisymmetric complex $\mathcal{N}=8$ central charge
matrix \cite{DFL}
\begin{equation}
Z_{AB}\left( \phi ,Q\right) =\left( Q^{T}L\left( \phi \right) \right)
_{AB}=\left( Q^{T}\right) _{\Lambda }L_{AB}^{\Lambda }\left( \phi \right) .
\label{central-charge-parameterization}
\end{equation}
$\phi $ denotes the $70$ real scalar fields parametrising the aforementioned
coset $\frac{E_{7(7)}}{SU(8)}$, $Q$ is the $\mathcal{N}=8$ charge vector
sitting in the fundamental irrepr. $\mathbf{56}$ of the $U$-duality group $%
E_{7\left( 7\right) }$. Moreover, $L_{AB}^{\Lambda }\left( \phi \right) $ is
the $\phi $-dependent coset representative, \textit{i.e.} a local section of
the principal bundle $E_{7\left( 7\right) }$ over $\frac{E_{7(7)}}{SU(8)}$
with structure group $SU(8)$.

The action of an element $g\in E_{7\left( 7\right) }$ on $V_{BH,\mathcal{N}%
=8}\left( \phi ,Q\right) $ is such that
\begin{equation}
V_{BH,\mathcal{N}=8}\left( \phi ,Q\right) =V_{BH,\mathcal{N}=8}\left( \phi
_{g},Q^{g}\right) =V_{BH,\mathcal{N}=8}\left( \phi _{g},\left( g^{-1}\right)
^{T}Q\right) ;
\end{equation}
thus, $V_{BH,\mathcal{N}=8}$ is not $E_{7\left( 7\right) }$-invariant,
because its coefficients (given by the components of $Q$) do not in general
remain the same. The situation changes if one restricts $g\equiv g_{Q}\in
H_{Q}$ to belong to the stabilizer $H_{Q}$ of one of the orbits $\frac{%
E_{7(7)}}{H_{Q}}$ spanned by the charge vector $Q$ within the $\mathbf{56}$
representation space of $E_{7\left( 7\right) }$ itself. In such a case:
\begin{equation}
Q^{g_{Q}}=Q\Rightarrow V_{BH,\mathcal{N}=8}\left( \phi ,Q\right) =V_{BH,%
\mathcal{N}=8}\left( \phi _{g_{Q}},Q\right) .  \label{11june}
\end{equation}
Then, it is natural to split the $70$ real scalar fields $\phi $ as $\phi
=\left\{ \phi _{Q},\breve{\phi}_{Q}\right\} $, where $\phi _{Q}\in \frac{%
H_{Q}}{\text{mcs}\left( H_{Q}\right) }\subsetneq \frac{E_{7\left( 7\right) }%
}{SU\left( 8\right) }$ and $\breve{\phi}_{Q}$ coordinatise the complement of
$\frac{H_{Q}}{\text{mcs}\left( H_{Q}\right) }$ in $\frac{E_{7\left( 7\right)
}}{SU\left( 8\right) }$. By denoting with
\begin{equation}
V_{BH,\mathcal{N}=8,crit}\left( \phi _{Q},Q\right) \equiv \left. V_{BH,%
\mathcal{N}=8}\left( \phi ,Q\right) \right| _{\frac{\partial V_{BH,\mathcal{N%
}=8}}{\partial \breve{\phi}_{Q}}=0}\left( \neq 0\right)
\end{equation}
the values of $V_{BH,\mathcal{N}=8}$ along the equations of motion for the
scalars $\breve{\phi}_{Q}$, the invariance of $V_{BH,\mathcal{N}%
=8,crit}\left( \phi _{Q},Q\right) $ under $H_{Q}$ directly follows from Eq. (%
\ref{11june}) :
\begin{equation}
V_{BH,\mathcal{N}=8,crit}\left( \left( \phi _{Q}\right) _{g_{Q}},Q\right)
=V_{BH,\mathcal{N}=8,crit}\left( \phi _{Q},Q\right) .
\end{equation}

Now, it is crucial to observe that $H_{Q}$ generally is a \textit{non-compact%
} Lie group; for instance, $H_{Q}=E_{6\left( 2\right) }\equiv \mathcal{H}_{%
\mathcal{N}=8}$ for $Q\in \mathcal{O}_{\frac{1}{8}-BPS,\mathcal{N}=8}$ given
by (\ref{N=8-BPS-large}), and $H_{Q}=E_{6\left( 6\right) }\equiv \widehat{%
\mathcal{H}}_{\mathcal{N}=8}$ for $Q\in \mathcal{O}_{nBPS,\mathcal{N}=8}$
given by (\ref{N=8-nBPS}). This implies $V_{BH,\mathcal{N}=8}$ to be
independent \textit{at its critical points} on the subset
\begin{equation}
\phi _{Q}\in \frac{H_{Q}}{\text{mcs}\left( H_{Q}\right) }\subsetneq \frac{%
E_{7\left( 7\right) }}{SU\left( 8\right) }.
\end{equation}
Thus, $\frac{H_{Q}}{\text{mcs}\left( H_{Q}\right) }$ can be regarded
as the ``moduli space'' of the attractor solutions supported by the
charge orbit $\frac{E_{7\left( 7\right) }}{H_{Q}}$. For
$\mathcal{N}=8$
non-degenerate critical points, supported by $\mathcal{O}_{\frac{1}{8}-BPS,%
\mathcal{N}=8}$ and $\mathcal{O}_{nBPS,\mathcal{N}=8}$, this
reasoning yields to the ``moduli spaces''
$\mathcal{M}_{\frac{1}{8}-BPS}$ and
$\mathcal{M}_{nBPS}$, respectively given by (\ref{N=8-BPS-large-mod}) and (%
\ref{N=8-nBPS}).

The results on $\mathcal{N}=8$ theory are summarized in the last row of
Tables 5 and 6.

The above arguments apply to a general, not necessarily supersymmetric,
Maxwell-Einstein theory with scalars coordinatising an homogeneous (not
necessarily symmetric) space. In particular, one can repeat the above
reasoning for all supergravities with $\mathcal{N}\geqslant 1$ based on
homogeneous (not necessarily symmetric) manifolds$\frac{G_{\mathcal{N}}}{H_{%
\mathcal{N}}}\equiv \frac{G_{\mathcal{N}}}{\text{mcs}\left( G_{\mathcal{N}%
}\right) }$, also in presence of matter multiplets. It is here worth
recalling that theories with $\mathcal{N}\geqslant 3$ all have symmetric
scalar manifolds (see \textit{e.g.} \cite{ADF-U-duality-revisited}).

A remarkable consequence is the existence of ``moduli spaces'' of
attractors is the following. By choosing $Q$ belonging to the orbit $\frac{%
G_{\mathcal{N}}}{H_{Q}}\subsetneq \mathbf{R}_{V}\left( G_{\mathcal{N}%
}\right) $ and supporting a class of non-degenerate critical points of $%
V_{BH,\mathcal{N}}$, \textit{up to some ``flat'' directions} (spanning the
``moduli space'' $\frac{H_{Q}}{\text{mcs}\left( H_{Q\;}\right) }\subsetneq
\frac{G_{\mathcal{N}}}{H_{\mathcal{N}}}$), \textit{all} such critical points
of $V_{BH,\mathcal{N}}$ in \textit{all} $\mathcal{N}\geqslant 0$
Maxwell-Einstein (super)gravities with an homogeneous (not necessarily
symmetric) scalar manifold (also in presence of matter multiplets) are
\textit{stable}, and thus they are \textit{attractors} in a generalized
sense. For $d=4$ supergravities, $H_{Q}=\mathcal{H}$, $\widehat{\mathcal{H}}$
or $\widetilde{\mathcal{H}}$ (see \textit{e.g.} Tables 5 and 6; see the
third, fifth and seventh Refs. of \cite{Revs}).

All this reasoning can be extended to a number of space-time dimensions $%
d\neq 4$ (see \textit{e.g.} \cite{Larsen-rev,AFMT-1,FMMS-1,CFMZ1-d=5}). As
found in \cite{GLS-1,stu-unveiled} for ``large'' charge orbits of $\mathcal{N%
}=2$, $d=4$ $stu$ model, and then proved in a model-independent way
in \cite {ADFT-FO-1}, the ``moduli spaces'' of charge orbits are
defined \textit{all along the corresponding scalar flows}, and thus
they can be interpreted as ``moduli spaces'' of unstabilized scalars
at the event horizon of the extremal black hole, as well as ``moduli
spaces'' of the ADM mass \cite{ADM} of the extremal black hole at
spatial infinity.

Remarkably, one can associate ``moduli spaces'' also to
non-attractive, ``small'' orbits, namely to charge orbits supporting
black hole configurations which have vanishing horizon area in the
Einsteinian approximation \cite{CFMZ1,Duff-N=8,CFMZ1-d=5}.
Differently from ``large'' orbits, for ``small'' orbits there exists
a ``moduli space'' also when the semi-simple part of $H_{Q}$ is
compact, and it has translational nature \cite {CFMZ1-d=5}. Clearly,
in the ``small'' case the interpretation at the event horizon breaks
down, simply because such an horizon does not exist at all,
\textit{at least} in Einsteinian supergravity approximation.

\section{``Moduli Spaces''of Attractors in $\mathcal{N}=2$, $d=4$\\Symmetric
Maxwell-Einstein Supergravities \label{Sect4}}

The arguments outlined in Sec. \ref{Sect3} can be used to determine the
``moduli spaces'' of non-BPS attractors (with $Z_{H}\neq 0$ or $Z_{H}=0$)
for all $\mathcal{N}=2$, $d=4$ Maxwell-Einstein supergravities with
symmetric scalar manifolds \cite{Ferrara-Marrani-2}.

After the fifth Ref. of \cite{AM-Refs}, it is known that, regardless of the
geometry of the vector multiplets' scalar manifold, the BPS non-degenerate
critical points of $V_{BH,\mathcal{N}=2}$ are \textit{stable}, and thus
define an attractor configuration in strict sense, in which all scalar
fields are stabilized in terms of charges by the Attractor Mechanism \cite
{AM-Refs}. This is ultimately due to the fact that the Hessian matrix $%
\mathbf{H}_{\frac{1}{2}-BPS}$ at such critical points has no massless modes
at all. Therefore, as as far as the metric of the scalar manifold is
non-singular and positive-definite and no massless degrees of freedom appear
in the theory, there is no ``moduli space'' for BPS attractors in \textit{any%
} $\mathcal{N}=2$, $d=4$ Maxwell-Einstein supergravity theory.

This is an important difference with respect to $\frac{1}{\mathcal{N}}$-BPS
attractors in $\mathcal{N}>2$-extended supergravities (see the third, fifth
and seventh Refs. of \cite{Revs}; for instance, in $\mathcal{N}=8$ theory $%
\frac{1}{8}$-BPS attractors exhibit the ``moduli space'' $\mathcal{M}_{\frac{%
1}{8}-BPS}$ given by (\ref{N=8-BPS-large-mod}). From a group
theoretical perspective, such a difference can be ascribed to the
\textit{compactness}
of the stabilizer $H_{0}$ of the ``large'' BPS charge orbit $\mathcal{O}_{%
\frac{1}{2}-BPS,\mathcal{N}=2}$ in the $\mathcal{N}=2$ symmetric case (see
Table 3). From a supersymmetry perspective, such a difference can be traced
back to the different degrees of supersymmetry preservation exhibited by
attractor solutions in theories with a different number $\mathcal{N}$ of
supercharges. Indeed, ($\frac{1}{2}$-)BPS attractors in theories with local $%
\mathcal{N}=2$ supersymmetry are \textit{maximally }supersymmetric (namely,
they preserve the \textit{maximum} number of supersymmetries out of the ones
related to the asymptotical Poincar\'{e} background). On the other hand, in $%
\mathcal{N}$-extended ($2<\mathcal{N}\leqslant 8$) supergravities BPS
attractors correspond to $\frac{1}{\mathcal{N}}$-BPS configurations, which
are are \textit{not maximally }supersymmetric. In these latter theories, th
maximally supersymmetric configurations correspond to vanishing black hole
entropy (at the two-derivative Einsteinian level).

Exploiting the observation below Eq. (\ref{N=8-nBPS}), it is possible to
determine the ``moduli spaces'' of non-BPS critical points ($Z_{H}\neq 0$ or
$Z_{H}=0$) of of $V_{BH,\mathcal{N}=2}$ for all $\mathcal{N}=2$, $d=4$
Maxwell-Einstein supergravities with symmetric scalar manifold. Consistent
with the notation introduced in Sec. \ref{Sect2} (recall (\ref{N=2-mcs's})),
the $\mathcal{N}=2$ non-BPS $Z_{H}\neq 0$ and $Z_{H}=0$ ``moduli spaces''
are respectively denoted by (see \cite{BFGM1,Ferrara-Marrani-2} for further
details on notation)
\begin{eqnarray}
\mathcal{M}_{nBPS,Z_{H}\neq 0} &=&\frac{\widehat{H}}{\text{mcs}\left(
\widehat{H}\right) }\equiv \frac{\widehat{H}}{\widehat{h}}; \\
\mathcal{M}_{nBPS,Z_{H}=0} &=&\frac{\widetilde{H}}{\text{mcs}\left(
\widetilde{H}\right) }\equiv \frac{\widetilde{H}}{\widetilde{h}}=\frac{%
\widetilde{H}}{\widetilde{h}^{\prime }\times U(1)}.
\end{eqnarray}
The results are reported in Tables 3 and 4 \cite{Ferrara-Marrani-2}.

\begin{table}[h!]
\par
\begin{center}
\begin{tabular}{|c||c|c|c|}
\hline & $
\begin{array}{c}
\\
\frac{\widehat{H}}{mcs(\widehat{H})} \\
~
\end{array}
$ & $
\begin{array}{c}
\\
r \\
~
\end{array}
$ & $
\begin{array}{c}
\\
$dim$_{\mathbb{R}} \\
~
\end{array}
$ \\ \hline $
\begin{array}{c}
\\
\mathbb{R}\oplus \mathbf{\Gamma }_{1,n-1},~n\in \mathbb{N} \\
~
\end{array}
$ & $SO(1,1)\times \frac{SO(1,n-1)}{SO(n-1)}~$ & $
\begin{array}{c}
\\
1~(n=1) \\
2~(n\geqslant 2)
\end{array}
~$ & $n$ \\ \hline $
\begin{array}{c}
\\
J_{3}^{\mathbb{O}} \\
~
\end{array}
$ & $\frac{E_{6(-26)}}{F_{4(-52)}}$ & $2$ & $26$ \\ \hline $
\begin{array}{c}
\\
J_{3}^{\mathbb{H}} \\
~
\end{array}
$ & $\frac{SU^{\ast }(6)}{USp(6)}~$ & $2$ & $14$ \\ \hline $
\begin{array}{c}
\\
J_{3}^{\mathbb{C}} \\
~
\end{array}
$ & $\frac{SL(3,\mathbb{C})}{SU(3)}$ & $2$ & $8~$ \\ \hline $
\begin{array}{c}
\\
J_{3}^{\mathbb{R}} \\
~
\end{array}
$ & $\frac{SL(3,\mathbb{R})}{SO(3)}$ & $2$ & $5$ \\ \hline
\end{tabular}
\end{center}
\caption{``\textbf{Moduli spaces'' of non-BPS }$Z_{H}\neq
0$\textbf{\
critical points of }$V_{BH,\mathcal{N}=2}$ \textbf{in }$\mathcal{N}$\textbf{$%
=2$, $d=4$ \textit{symmetric} Maxwell-Einstein supergravities. They are the }%
$\mathcal{N}$$=2$\textbf{,} $d=5$ \textbf{symmetric real special
manifolds}}
\end{table}

\begin{table}[h!]
\par
\begin{center}
\begin{tabular}{|c||c|c|c|}
\hline & $
\begin{array}{c}
\\
\frac{\widetilde{H}}{mcs(\widetilde{H})}\equiv\frac{\widetilde{H}}{\widetilde{h}%
^{\prime }\times U(1)} \\
~
\end{array}
$ & $
\begin{array}{c}
\\
r \\
~
\end{array}
$ & $
\begin{array}{c}
\\
$dim$_{\mathbb{C}} \\
~
\end{array}
$ \\ \hline\hline $
\begin{array}{c}
\text{\textit{minimal~coupling}} \\
n\in \mathbb{N}
\end{array}
$ & $\frac{SU(1,n-1)}{U(1)\times SU(n-1)}~$ & $1$ & $n~-1$ \\ \hline
$
\begin{array}{c}
\\
\mathbb{R}\oplus \mathbf{\Gamma }_{1,n-1},~n\in \mathbb{N} \\
~
\end{array}
$ & $\frac{SO(2,n-2)}{SO(2)\times SO(n-2)},~n\geqslant 3~$ & $
\begin{array}{c}
\\
1~(n=3) \\
2~(n\geqslant 4)
\end{array}
~$ & $n-2$ \\ \hline $
\begin{array}{c}
\\
J_{3}^{\mathbb{O}} \\
~
\end{array}
$ & $\frac{E_{6(-14)}}{SO(10)\times U(1)}$ & $2$ & $16$ \\ \hline $
\begin{array}{c}
\\
J_{3}^{\mathbb{H}} \\
~
\end{array}
$ & $\frac{SU(4,2)}{SU(4)\times SU(2)\times U(1)}~$ & $2$ & $8$ \\
\hline $
\begin{array}{c}
\\
J_{3}^{\mathbb{C}} \\
~
\end{array}
$ & $\frac{SU(2,1)}{SU(2)\times U(1)}\times
\frac{SU(1,2)}{SU(2)\times U(1)}$ & $2$ & $4$ \\ \hline $
\begin{array}{c}
\\
J_{3}^{\mathbb{R}} \\
~
\end{array}
$ & $\frac{SU(2,1)}{SU(2)\times U(1)}$ & $1$ & $2$ \\ \hline
\end{tabular}
\end{center}
\caption{``\textbf{Moduli spaces'' of non-BPS }$Z_{H}=0$\textbf{\
critical
points of }$V_{BH,\mathcal{N}=2}$ \textbf{in }$\mathcal{N}$\textbf{$=2$, $%
d=4 $ \textit{symmetric} Maxwell-Einstein supergravities. They are
(non-special) symmetric K\"{a}hler manifolds}}
\end{table}

As observed below Eq. (\ref{N=8-nBPS-mod}), the non-BPS $Z_{H}\neq 0$
``moduli spaces'' are nothing but the scalar manifolds of minimal ($\mathcal{%
N}=2$) Maxwell-Einstein supergravity in $d=5$ space-time dimensions \cite
{GST}. Their real dimension dim$_{\mathbb{R}}$ (rank $r$) is the complex
dimension dim$_{\mathbb{C}}$ (rank $r$) of the $\mathcal{N}=2$, $d=4$
symmetric special K\"{a}hler manifolds listed in Table 1, minus one. With
the exception of the $n=1$ element of the generic Jordan family $\mathbb{R}%
\oplus \mathbf{\Gamma }_{1,n-1}$ (the so-called $st^{2}$ model) having $%
\frac{\widehat{H}}{\widehat{h}}=SO(1,1)$ with rank $r=1$, all non-BPS $%
Z_{H}\neq 0$ ``moduli spaces'' have rank $r=2$. The results reported in
Table 3 are consistent with the ``$n_{V}+1$ / $n_{V}-1$'' mass degeneracy
splitting of non-BPS $Z_{H}\neq 0$ attractors \cite
{TT,BFGM1,TT2,Ferrara-Marrani-1}, holding for a generic special K\"{a}hler
cubic geometry of complex dimension $n_{V}$.

The non-BPS $Z_{H}=0$ ``moduli spaces'', reported in Table 4, are symmetric
(generally non-special) K\"{a}hler manifolds. Note that in the $n=1$ and $%
n=2 $ elements of the generic Jordan family $\mathbb{R}\oplus \mathbf{\Gamma
}_{1,n-1}$ (the so-called $st^{2}$ and $stu$ models, respectively), there
are no non-BPS $Z_{H}=0$ ``flat'' directions at all (see Appendix II of \cite
{BFGM1}, and \cite{Ferrara-Marrani-2}). By recalling the definition $A\equiv
dim_{\mathbb{R}}\mathbb{A}$ given above, the results reported in Table 4
\cite{Ferrara-Marrani-2} imply that the the non-BPS $Z_{H}=0$ ``moduli
spaces'' of $\mathcal{N}=2$, $d=4$ ``magic'' supergravities supergravities
have complex dimension $2A$. As observed in \cite{Ferrara-Marrani-2}, the
non-BPS $Z_{H}=0$ ``moduli space'' of $\mathcal{N}=2$, $d=4$ ``magic''
supergravity associated to $J_{3}^{\mathbb{O}}$ is the manifold $\frac{%
E_{6(-14)}}{SO(10)\otimes U(1)}$, which is related to another exceptional
Jordan triple system over $\mathbb{O}$, as found long time ago in \cite{GST}.

\begin{table}[p]
\begin{center}
\begin{tabular}{|c||c|c|c|}
\hline & $
\begin{array}{c}
\\
\frac{1}{\mathcal{N}}\text{-BPS orb }\frac{G_{\mathcal{N}}}{\mathcal{H}_{%
\mathcal{N}}} \\
~
\end{array}
$ & $
\begin{array}{c}
\\
\text{nBPS }Z_{AB,H}\neq 0\text{ orb}~\frac{G_{\mathcal{N}}}{\widehat{%
\mathcal{H}}_{\mathcal{N}}} \\
~
\end{array}
$ & $
\begin{array}{c}
\\
\text{nBPS }Z_{AB,H}=0\text{ orb }\frac{G_{\mathcal{N}}}{\widetilde{\mathcal{%
H}}_{\mathcal{N}}} \\
\\
~
\end{array}
$ \\ \hline\hline $
\begin{array}{c}
\\
\mathcal{N}=3 \\
n\in \mathbb{N}~
\end{array}
$ & $\frac{SU(3,n)}{SU(2,n)}~$ & $-$ & $\frac{SU(3,n)}{SU(3,n-1)}~$
\\ \hline $
\begin{array}{c}
\\
\mathcal{N}=4 \\
~n\in \mathbb{N},~\mathbb{R}\oplus \mathbf{\Gamma }_{5,n-1}
\end{array}
$ & $\frac{SL(2,\mathbb{R})}{SO\left( 2\right) }\times \frac{SO(6,n)}{SO(4,n)}~$ & $%
\frac{SL(2,\mathbb{R})}{SO\left( 1,1\right) }\times \frac{SO(6,n)}{SO(5,n-1)}~$ & $%
\frac{SL(2,\mathbb{R})}{SO\left( 2\right) }\times
\frac{SO(6,n)}{SO(6,n-2)}$
\\ \hline $
\begin{array}{c}
\\
\mathcal{N}=5 \\
~M_{1,2}\left( \mathbb{O}\right)
\end{array}
$ & $\frac{SU(1,5)}{SU(3)\times SU\left( 2,1\right) }$ & $-$ & $-$
\\ \hline $
\begin{array}{c}
\\
\mathcal{N}=6 \\
~J_{3}^{\mathbb{H}}
\end{array}
$ & $\frac{SO^{\ast }(12)}{SU(4,2)}~$ & $\frac{SO^{\ast }(12)}{SU^{\ast }(6)}%
~$ & $\frac{SO^{\ast }(12)}{SU(6)}~$ \\ \hline $
\begin{array}{c}
\\
\mathcal{N}=8 \\
~J_{3}^{\mathbb{O}_{s}}
\end{array}
$ & $\frac{E_{7\left( 7\right) }}{E_{6\left( 2\right) }}$ & $\frac{%
E_{7\left( 7\right) }}{E_{6\left( 6\right) }}$ & $-~$ \\ \hline
\end{tabular}
\end{center}
\caption{\textbf{Charge orbits supporting extremal black hole attractors in }%
$\mathcal{N}$\textbf{$\geqslant 3$-extended, $d=4$ supergravities} \textbf{(}%
$n$\textbf{\ is the number of matter multiplets) (see the fifth Ref.
of
\protect\cite{Revs}).} \textbf{The related Euclidean degree-}$\mathbf{3}$%
\textbf{\ Jordan algebra\ is also given (\textit{if any}).
$M_{1,2}\left( \mathbb{O}\right) $ is the Jordan triple system (not
upliftable to $d=5$) generated by $2\times 1$ Hermitian matrices
over $\mathbb{O}$ \protect\cite {GST}.}}
\end{table}

\begin{table}[p]
\begin{center}
\begin{tabular}{|c||c|c|c|}
\hline & $
\begin{array}{c}
\\
\frac{1}{\mathcal{N}}\text{-BPS} \\
\text{``moduli space'' }\frac{\mathcal{H}_{\mathcal{N}}}{mcs(\mathcal{H}_{\mathcal{N}})}\text{ } \\
~
\end{array}
$ & $
\begin{array}{c}
\\
\text{nBPS }Z_{AB,H}\neq 0 \\
\text{``moduli space'' }\frac{\widehat{\mathcal{H}}_{\mathcal{N}}}{mcs(\widehat{\mathcal{H}}_{\mathcal{N}})} \\
~
\end{array}
$ & $
\begin{array}{c}
\\
\text{nBPS }Z_{AB,H}=0 \\
\text{``moduli space'' }\frac{\widetilde{\mathcal{H}}_{\mathcal{N}}}{mcs(\widetilde{\mathcal{H}}_{\mathcal{N}})} \\
~
\end{array}
$ \\ \hline\hline $
\begin{array}{c}
\\
\mathcal{N}=3 \\
~
\end{array}
$ & $\frac{SU(2,n)}{SU(2)\times SU\left( n\right) \times U\left(
1\right) }~$ & $-$ & $\frac{SU(3,n-1)}{SU(3)\times SU\left(
n-1\right) \times U\left( 1\right) }~$ \\ \hline $
\begin{array}{c}
\\
\mathcal{N}=4 \\
~
\end{array}
$ & $\frac{SO(4,n)}{SO(4)\times SO\left( n\right) }~$ & $SO(1,1)\times \frac{%
SO(5,n-1)}{SO(5)\times SO\left( n-1\right) }~$ & $\frac{SO(6,n-2)}{%
SO(6)\times SO\left( n-2\right) }$ \\ \hline $
\begin{array}{c}
\\
\mathcal{N}=5 \\
~
\end{array}
$ & $\frac{SU\left( 2,1\right) }{SU\left( 2\right) \times U\left(
1\right) }$ & $-$ & $-$ \\ \hline $
\begin{array}{c}
\\
\mathcal{N}=6 \\
~
\end{array}
$ & $\frac{SU(4,2)}{SU(4)\times SU\left( 2\right) \times U\left(
1\right) }~$ & $\frac{SU^{\ast }(6)}{USp\left( 6\right) }~$ & $-$ \\
\hline $
\begin{array}{c}
\\
\mathcal{N}=8 \\
~
\end{array}
$ & $\frac{E_{6\left( 2\right) }}{SU\left( 6\right) \times SU\left(
2\right) }$ & $\frac{E_{6\left( 6\right) }}{USp\left( 8\right) }$ &
$-~$ \\ \hline
\end{tabular}
\end{center}
\caption{``\textbf{Moduli spaces'' of black hole attractor solutions in }$%
\mathcal{N}$\textbf{$\geqslant 3$-extended, $d=4$ supergravities.
}$n$ \textbf{is the number of matter multiplets} (see the fifth Ref.
of \protect\cite {Revs})}
\end{table}

\setcounter{equation}0

\section{\label{N>2,d=4}$\mathcal{N}\geqslant 3$-Extended, $d=4$
Supergravities}

As anticipated above, the scalar manifolds of all $d=4$ supergravity
theories with $\mathcal{N}\geqslant 3$ supercharges are symmetric
spaces (they are reported \textit{e.g.} in Table 6 of \cite{LA08}).
Both $\frac{1}{\mathcal{N}}$-BPS and non-BPS attractors exhibit a
related ``moduli space''. An example is provided by the maximal
theory,
already reviewed in Sec. \ref{Sect3}. As mentioned above, the \textit{%
non-compactness} of the stabilizer group of the corresponding supporting
charge orbit is the ultimate reason of the existence of the ``moduli
spaces'' of attractor solutions \cite{Ferrara-Marrani-1,Ferrara-Marrani-2}
(see also the fifth Ref. of \cite{Revs}).

By performing a supersymmetry truncation down to $\mathcal{N}=2$ \cite
{ADF-SUSY-trunc-1,ADF-U-duality-d=4,Ferrara-Marrani-1}, the $\frac{1}{%
\mathcal{N}}$-BPS ``flat'' directions of $V_{BH,\mathcal{N}\text{ }}$can be
interpreted in terms of left-over $\mathcal{N}=2$ hypermultiplets' scalar
degrees of freedom. As studied in \cite{Ferrara-Marrani-1}, for non-BPS
``flat'' directions the situation is more involved, and an easy
interpretation in terms of truncated-away hypermultiplets' scalars degrees
of freedom is generally lost.

Tables 5 and 6 report all classes of charge orbits supporting attractor
solutions in $\mathcal{N}\geqslant 3$-extended supergravity theories in $d=4$
space-time dimensions (see the third, fifth and seventh Refs. of \cite{Revs}%
).

\section*{\textbf{Acknowledgments}}

The contents of this brief report result from collaborations with
Stefano Bellucci, Murat G\"{u}naydin, Renata Kallosh, and especially
Sergio Ferrara, which are gratefully acknowledged.

\end{document}